\documentclass[sigconf,nonacm]{acmart}
\setcopyright{none}
\usepackage{amsmath,graphicx}
\usepackage{tikz,pgfplots}
\pgfplotsset{compat=1.17}
\usetikzlibrary{patterns,calc,arrows.meta,positioning,shapes.geometric}
\usepackage{xurl}
\definecolor{FDdark}{RGB}{27,67,150}
\definecolor{FDlight}{RGB}{115,145,192}
\definecolor{FDblue}{RGB}{46,91,161}
\colorlet{Blue}{FDblue}
\definecolor{Warm}{HTML}{A85C4F}
\colorlet{Pale}{FDlight!55!white}
\definecolor{Ink}{HTML}{222C3B}
\colorlet{Rule}{FDlight!24!white}
\definecolor{Rose}{HTML}{BE8378}
\graphicspath{{assets/}}
\tikzset{
 pics/person/.style args={#1}{code={
  \filldraw[fill=#1!13,draw=#1,line width=.7pt] (0,.28) circle (.17);
  \filldraw[fill=#1!13,draw=#1,line width=.7pt,rounded corners=3pt]
   (-.28,-.35)--(-.28,-.10) .. controls (-.25,.09) and (.25,.09) .. (.28,-.10)--(.28,-.35)--cycle;
 }},
 pics/page/.style args={#1}{code={
  \draw[draw=#1,fill=white,line width=.7pt] (-.24,-.33)--(.24,-.33)--(.24,.21)--(.12,.33)--(-.24,.33)--cycle;
  \draw[#1,line width=.55pt] (.12,.33)--(.12,.21)--(.24,.21);
  \foreach \y in {-.18,-.06,.06}{\draw[#1!70,line width=.65pt] (-.14,\y)--(.12,\y);}
 }},
 pics/terminal/.style args={#1}{code={
  \draw[draw=#1,fill=white,line width=.8pt,rounded corners=2pt] (-.47,-.29) rectangle (.47,.32);
  \fill[#1!15] (-.46,.18) rectangle (.46,.30);
  \draw[#1,line width=.8pt] (-.22,.07)--(-.32,-.02)--(-.22,-.11) (.22,.07)--(.32,-.02)--(.22,-.11) (.07,.09)--(-.03,-.14);
  \draw[#1,line width=.75pt] (0,-.30)--(0,-.40) (-.22,-.40)--(.22,-.40);
 }},
 pics/chip/.style args={#1}{code={
  \draw[#1,fill=#1!7,line width=.75pt] (-.33,-.27) rectangle (.33,.27);
  \foreach \x in {-.22,0,.22}{\draw[#1] (\x,.27)--(\x,.40) (\x,-.27)--(\x,-.40);}
  \foreach \y in {-.16,0,.16}{\draw[#1] (-.33,\y)--(-.46,\y) (.33,\y)--(.46,\y);}
  \fill[#1!50] (-.22,-.15) rectangle (-.02,.13);
  \fill[#1!25] (.04,-.15) rectangle (.23,-.03);
  \fill[#1!35] (.04,.03) rectangle (.23,.14);
 }}
}
\hypersetup{colorlinks=true,linkcolor=Ink,citecolor=Blue,urlcolor=Blue,pdftitle={When Algorithmic Exploration Becomes Cheap: A Case Study of Agentic Research in EDA},pdfauthor={Keren Zhu; Yu Deng; Xiaoyu Hao; Liwen Jiang; Zijian Jiang; Cunqing Lan; Boxiang Song; Pujun Su; Yaojia Wang},pdfsubject={Agentic algorithm research in electronic design automation}}
\color{Ink}
\pgfplotsset{every axis/.append style={font=\sffamily\figtext,text=Ink,axis line style={Ink,thin},tick style={Ink,thin},tick label style={font=\sffamily\figsmall,text=Ink},label style={font=\sffamily\figtext},grid style={Rule,thin},legend style={draw=none,fill=none,font=\sffamily\figsmall},enlargelimits=false}}
\newcommand{\panel}[1]{\textsf{\textbf{#1}}}
\newenvironment{keypoints}{\begin{itemize}\setlength{\itemsep}{3pt}\setlength{\parsep}{0pt}\setlength{\parskip}{0pt}}{\end{itemize}}
\newcommand{\figsmall}{\fontsize{8.5}{10}\selectfont}
\newcommand{\figtext}{\fontsize{9}{11}\selectfont}
\newcommand{\figtitle}{\fontsize{10}{12}\selectfont}
\providecommand{\doi}[1]{\href{https://doi.org/#1}{\nolinkurl{doi:#1}}}
\newcommand{\arxiv}[1]{\href{https://arxiv.org/abs/#1}{arXiv:#1}}
\title[When Algorithmic Exploration Becomes Cheap]{When Algorithmic Exploration Becomes Cheap:\\A Case Study of Agentic Research in EDA}
\author{Keren Zhu}
\email{krzhu@fudan.edu.cn}
\author{Yu Deng}
\authornote{Alphabetical order by surname.}
\author{Xiaoyu Hao}
\authornotemark[1]
\author{Liwen Jiang}
\authornotemark[1]
\author{Zijian Jiang}
\authornotemark[1]
\author{Cunqing Lan}
\authornotemark[1]
\author{Boxiang Song}
\authornotemark[1]
\author{Pujun Su}
\authornotemark[1]
\author{Yaojia Wang}
\authornotemark[1]
\affiliation{\department{College of Integrated Circuits and Micro-Nano Electronics}\institution{Fudan University}\city{Shanghai}\country{China}}
\renewcommand{\shortauthors}{Zhu et al.}
\makeatletter
\def\@mkauthors{%
 \global\setbox\mktitle@bx=\vbox{%
  \box\mktitle@bx
  \centering
  {\fontsize{11}{15}\selectfont
   Keren Zhu\quad Yu Deng\textsuperscript{*}\quad Xiaoyu Hao\textsuperscript{*}\quad Liwen Jiang\textsuperscript{*}\quad Zijian Jiang\textsuperscript{*}\\
   Cunqing Lan\textsuperscript{*}\quad Boxiang Song\textsuperscript{*}\quad Pujun Su\textsuperscript{*}\quad Yaojia Wang\textsuperscript{*}\par}
  \vspace{3pt}
  {\fontsize{9}{12}\selectfont College of Integrated Circuits and Micro-Nano Electronics\\
   Fudan University, Shanghai, China\\
   \href{mailto:krzhu@fudan.edu.cn}{krzhu@fudan.edu.cn}\par}
  \vspace{10pt}}}
\makeatother
\begin{document}
\begin{abstract}
As EDA researchers, we conducted eight deliberate trials of agentic algorithm exploration, selecting several topics outside our areas of depth. One faculty member and seven students participated, including students without publication experience. With limited intervention in the algorithms, agents developed mathematical constructions, analyzed existing tools, and implemented improvements; some efforts fell short of their practical goals. We also used AI to collect, classify, and analyze 8,420 papers from four EDA conferences and two journals over 2022--2026. Among 2,380 primary-core papers, we classified 97.7\% from titles and abstracts as computationally closed, including work on new formulations. Together, these observations suggest that much of EDA offers an executable environment for increasingly accessible algorithm research. We see an opportunity for tool developers to investigate ideas they previously lacked time to pursue. We also ask how EDA should validate and reward research when results become easier to produce than to examine, and what papers and venue labels will continue to tell us about a contribution.
\end{abstract}


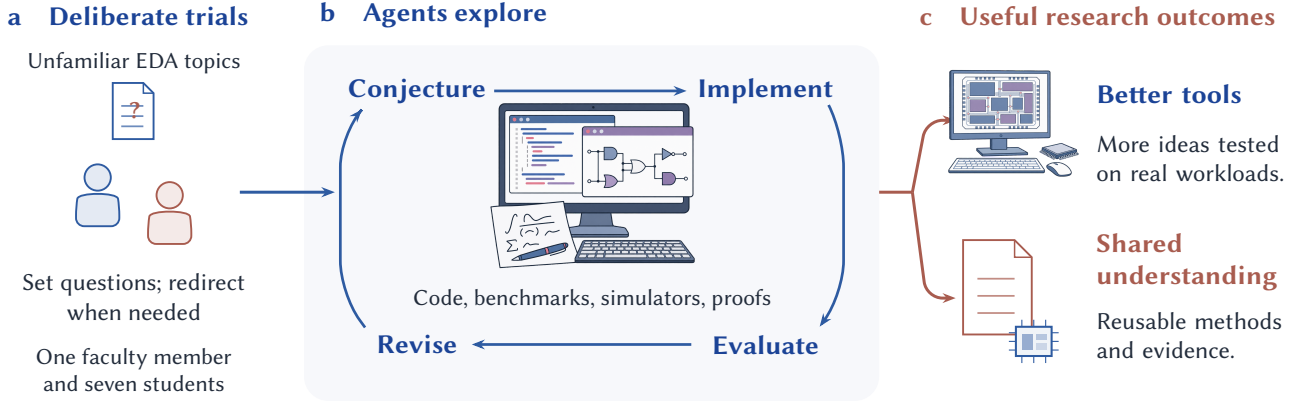
\begin{teaserfigure}
\centering
\resizebox{\textwidth}{!}{%
\begin{tikzpicture}[x=1cm,y=1cm,font=\sffamily\figtext,text=Ink]
\path[use as bounding box] (0,.48) rectangle (17.4,5.95);
\node[anchor=west,font=\sffamily\bfseries\figtitle,text=FDdark] at (.08,5.64) {a\quad Deliberate trials};
\node[anchor=west,font=\sffamily\bfseries\figtitle,text=FDdark] at (4.12,5.64) {b\quad Agents explore};
\node[anchor=west,font=\sffamily\bfseries\figtitle,text=Warm] at (11.88,5.64) {c\quad Useful research outcomes};
\fill[Blue!4,rounded corners=9pt] (4.03,.65) rectangle (11.47,5.25);
\begin{scope}[shift={(1.82,3.26)},scale=1.35]
 \pic at (-.34,0) {person=Blue};
 \pic at (.35,-.17) {person=Warm};
\end{scope}
\pic[scale=1.1] at (1.82,4.40) {page=Blue};
\node[font=\sffamily\bfseries\figtext,text=Warm] at (1.84,4.41) {?};
\node[font=\sffamily\figsmall] at (1.83,5.02) {Unfamiliar EDA topics};
\node[align=center] at (1.83,2.00) {Set questions; redirect\\when needed};
\node[align=center,font=\sffamily\figsmall] at (1.83,1.05) {One faculty member\\and seven students};
\draw[Blue,line width=1.1pt,-{Stealth[length=5pt]}] (3.20,3.35)--(4.43,3.35);
\node[inner sep=0] at (7.76,3.43) {\includegraphics[width=2.88cm]{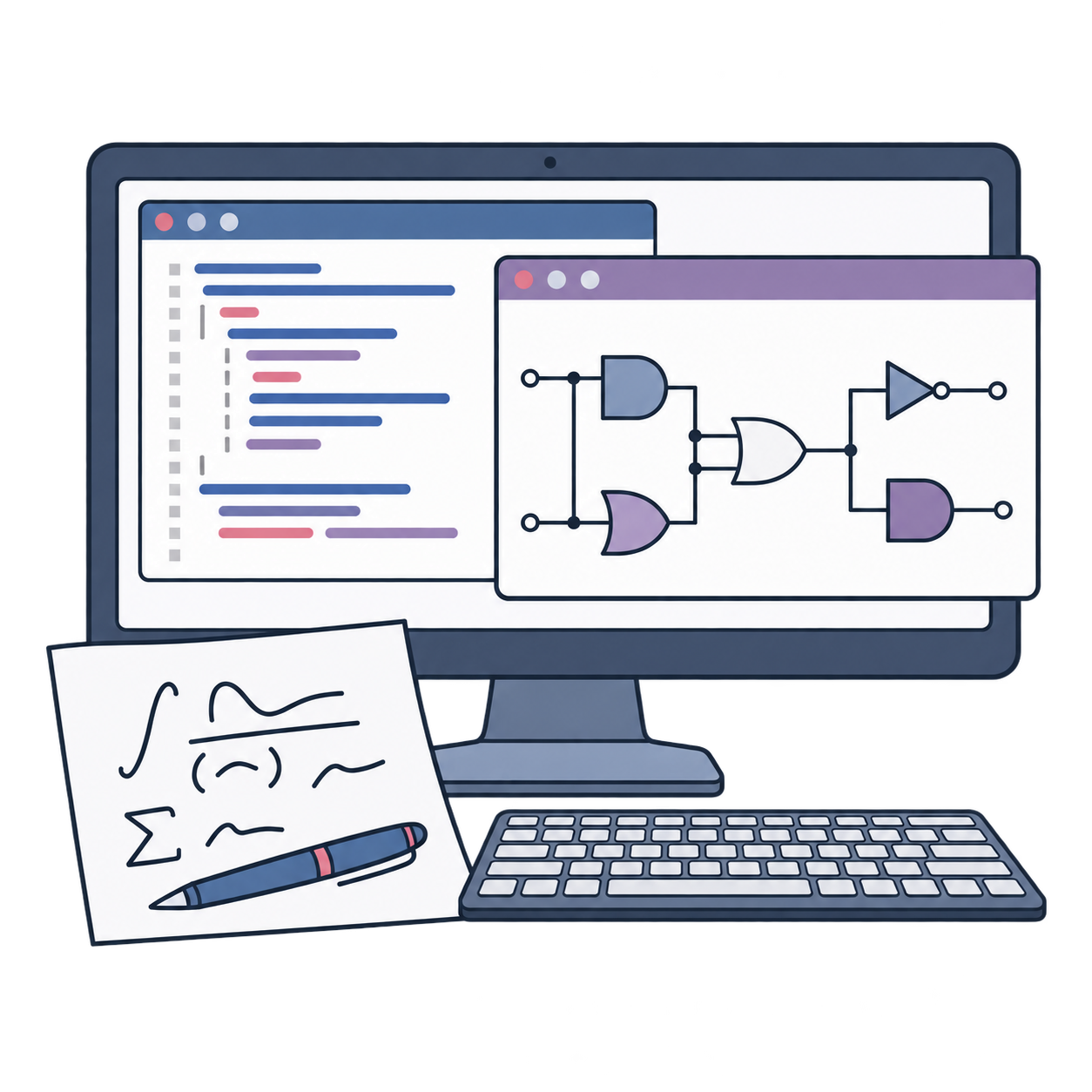}};
\node[fill=Blue!4,inner sep=3pt,font=\sffamily\bfseries\figtitle,text=FDdark] at (5.49,4.66) {Conjecture};
\node[fill=Blue!4,inner sep=3pt,font=\sffamily\bfseries\figtitle,text=FDdark] at (10.00,4.66) {Implement};
\node[fill=Blue!4,inner sep=3pt,font=\sffamily\bfseries\figtitle,text=FDdark] at (10.00,1.38) {Evaluate};
\node[fill=Blue!4,inner sep=3pt,font=\sffamily\bfseries\figtitle,text=FDdark] at (5.49,1.38) {Revise};
\draw[Blue,line width=1.0pt,-{Stealth[length=4.8pt]}] (6.49,4.66)--(8.98,4.66);
\draw[Blue,line width=1.0pt,-{Stealth[length=4.8pt]},rounded corners=12pt]
 (10.82,4.47)--(11.01,4.10)--(11.01,2.04)--(10.70,1.61);
\draw[Blue,line width=1.0pt,-{Stealth[length=4.8pt]}] (9.04,1.38)--(6.20,1.38);
\draw[Blue,line width=1.0pt,-{Stealth[length=4.8pt]},rounded corners=12pt]
 (4.82,1.60)--(4.50,2.04)--(4.50,4.10)--(4.78,4.42);
\node[font=\sffamily\figsmall] at (7.76,1.96) {Code, benchmarks, simulators, proofs};
\draw[Warm,line width=1.1pt] (11.47,3.35)--(11.90,3.35);
\draw[Warm,line width=1.1pt,-{Stealth[length=4.8pt]},rounded corners=6pt] (11.90,3.35)--(12.08,4.28)--(12.46,4.28);
\draw[Warm,line width=1.1pt,-{Stealth[length=4.8pt]},rounded corners=6pt] (11.90,3.35)--(12.08,1.98)--(12.46,1.98);
\node[inner sep=0] at (13.17,4.26) {\includegraphics[width=1.92cm]{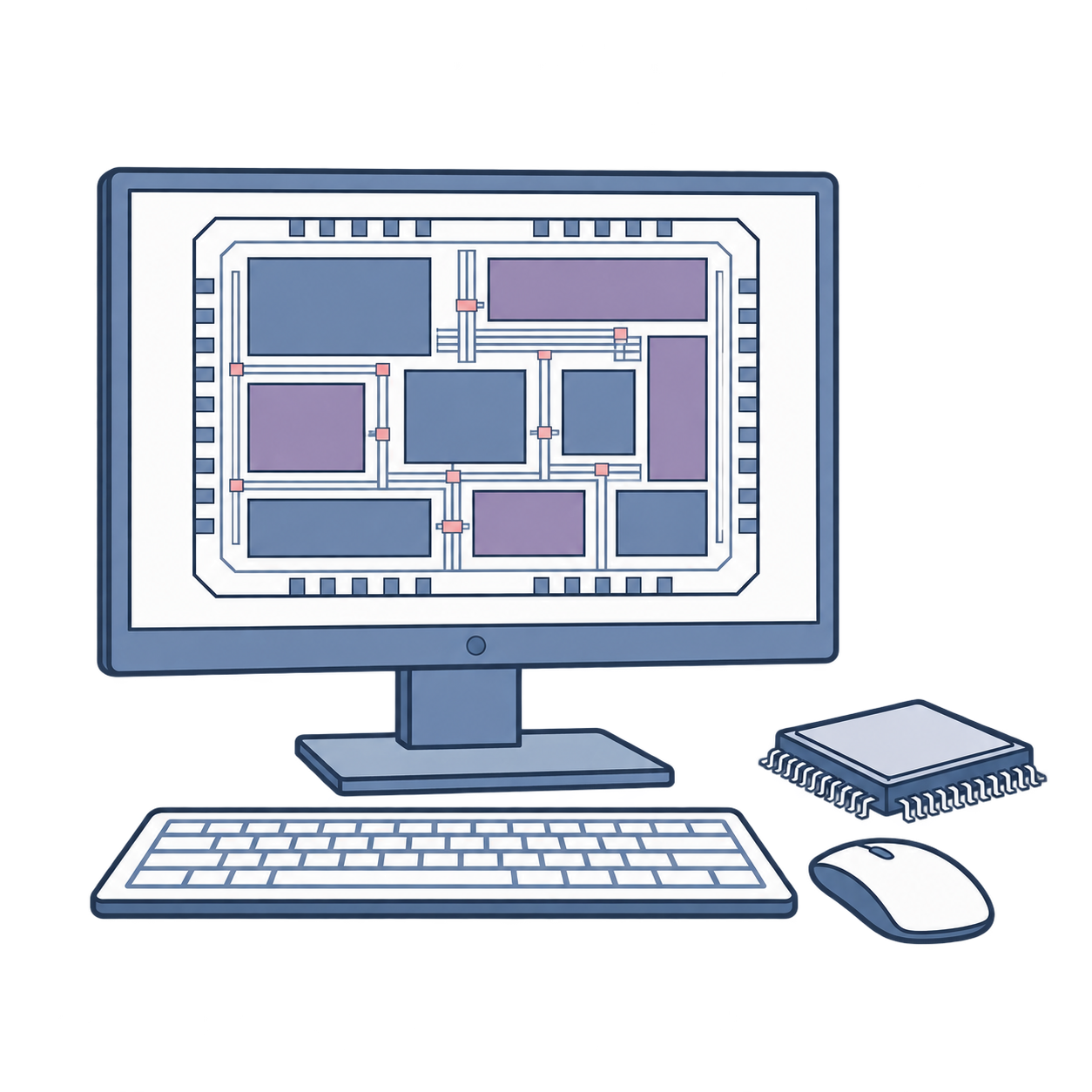}};
\node[anchor=west,align=left,text width=2.8cm,font=\sffamily\bfseries\figtitle,text=FDdark] at (14.17,4.62) {Better tools};
\node[anchor=west,align=left,text width=2.8cm] at (14.17,3.79) {More ideas tested\\on real workloads.};
\pic[scale=1.85] at (13.00,2.12) {page=Warm};
\pic[scale=.70] at (13.47,1.42) {chip=Blue};
\node[anchor=west,align=left,text width=2.8cm,font=\sffamily\bfseries\figtitle,text=Warm] at (14.17,2.45) {Shared\\understanding};
\node[anchor=west,align=left,text width=2.8cm] at (14.17,1.50) {Reusable methods\\and evidence.};
\end{tikzpicture}}
\caption{A deliberate set of EDA research trials. We selected unfamiliar questions and delegated much of the algorithmic exploration. The outcomes are considered through their contribution to tools and transferable understanding.}
\Description{A faculty member and students choose unfamiliar EDA questions. An executable research loop connects to an improved EDA tool and a reusable research result.}
\label{fig:perspective}
\end{teaserfigure}
\maketitle

\section{Why we ran these trials}

We were inspired by recent breakthroughs in AI-assisted mathematics, including new mathematical constructions and progress on research-level proof problems\cite{alphamath,firstproof,mathnews}. As EDA researchers, we wanted to find out how far similar exploration could go in our field. We organized eight deliberate Agentic Research Trials (ART), selecting several topics in which we had little prior depth. One faculty member and seven students participated. Some students had neither published a paper nor completed a research project from conception through publication. We specified demanding research questions and sought to interfere as little as possible with the algorithms used to answer them.

The trials covered rooted rectilinear spanning trees, logic optimization, global placement, hypergraph contraction, clock-tree realizability, timing-aware remapping, timing-aware rewriting, and LUT-based FPGA optimization. We asked agents to construct guarantees, challenge conjectures, identify useful structure, or turn an idea into an implementation. We sometimes changed a question, supplied a direction, or rejected an inadequate comparison. We retained unsuccessful branches and unmet goals alongside the resulting constructions and tools.

\paragraph{Three questions motivated the trials.}
\begin{keypoints}
\item How far can a small EDA group explore an unfamiliar topic without supplying the algorithmic route?
\item Which results transfer from a proof or executable experiment into a useful tool?
\item How should EDA evaluate and recognize research as this exploration becomes less expensive?
\end{keypoints}

The experiences changed our expectations about the specialist effort needed to begin an investigation. They also exposed a persistent gap between obtaining a technically valid result and improving a mature tool. Our interest is in using the new capacity to strengthen EDA research, with particular attention to how results transfer into real design tools (Fig.~\ref{fig:perspective}).

Work in EDA already spans conversational hardware design, domain-adapted assistants, and autonomous tool orchestration\cite{chipchat,chipnemo,chateda,llmeda}. ART extends the question to developing the algorithms inside a tool. The wider history of AI-assisted discovery\cite{wangscience} includes mathematical conjectures\cite{davies}, matrix multiplication\cite{alphatensor}, and sorting routines integrated into LLVM\cite{alphadev}. FunSearch and AlphaEvolve use executable feedback for discovery\cite{funsearch,alphaevolve}; The AI Scientist explores end-to-end research production\cite{aiscientist}. We paired our trials with an analysis of EDA's publication record, asking how much research already has a computationally closed loop and what changes appear in its distribution.

\section{How we organized the research trials}

ART used sustained pressure and independent judgment to push agents beyond their first plausible answer. We set a demanding endpoint, judged each candidate against it, and sent inadequate results back for further work. Participants used different prompts, tools, and divisions of labor. The shared approach was to constrain what would count as a result while leaving the algorithmic route open (Fig.~\ref{fig:art-method}).

\begin{figure*}[t]
\centering
\resizebox{\textwidth}{!}{%
\begin{tikzpicture}[x=1cm,y=1cm,font=\sffamily\figtext,text=Ink]
\path[use as bounding box] (0,.02) rectangle (17.4,6.48);
\node[anchor=west,font=\sffamily\bfseries\figtitle,text=FDdark] at (.10,6.15) {a\quad Open routes; a fixed acceptance gate};
\node[anchor=west,font=\sffamily\bfseries\figtitle,text=FDdark] at (11.05,6.15) {b\quad Tighten the demands};
\draw[Rule] (10.70,.25)--(10.70,6.20);
\fill[Blue!4,rounded corners=9pt] (.10,1.98) rectangle (6.47,5.70);
\node[anchor=west,font=\sffamily\figtext,text=FDdark] at (.35,5.36) {Presume the goal is attainable.};
\node[anchor=west,font=\sffamily\figtext,text=FDdark] at (.35,4.96) {Presume prior routes are insufficient.};
\draw[Ink!65,densely dashed,line width=1.05pt]
 (.67,3.39).. controls (1.45,3.42) and (1.71,4.28)..(2.83,4.28)--(5.76,4.28);
\node[fill=Blue!4,inner sep=2pt,font=\sffamily\figsmall] at (3.55,4.30) {Familiar route};
\draw[Warm,line width=1.3pt] (5.65,4.17)--(5.87,4.39) (5.65,4.39)--(5.87,4.17);
\draw[Blue,line width=1.5pt]
 (.67,3.39).. controls (1.66,4.10) and (2.30,2.95)..(3.10,3.38)
 .. controls (4.20,4.00) and (4.65,3.50)..(6.26,3.50);
\draw[Warm,line width=1.2pt]
 (.67,3.39).. controls (1.52,2.18) and (2.52,2.16)..(3.45,2.71)
 .. controls (4.59,3.40) and (5.82,2.88)..(7.62,2.88);
\filldraw[fill=white,draw=Blue,line width=1.2pt] (.67,3.39) circle (.10);
\node[font=\sffamily\figtext] at (3.34,2.09) {Explore materially different approaches};
\pic[scale=1.16] at (7.83,5.22) {terminal=Blue};
\node[anchor=west,align=left,font=\sffamily\figtext] at (8.63,5.20) {Independent\\evaluator};
\draw[Blue,line width=.95pt,-{Stealth[length=4.5pt]}] (7.83,4.67)--(7.83,4.27);
\draw[Ink,line width=1.6pt] (7.83,2.37)--(7.83,3.22) (7.83,3.78)--(7.83,4.18);
\draw[Ink,line width=1.0pt] (7.58,2.37)--(8.08,2.37) (7.58,4.18)--(8.08,4.18);
\draw[Ink!65,line width=.7pt] (8.03,2.37)--(8.03,3.22) (8.03,3.78)--(8.03,4.18);
\node[font=\sffamily\bfseries\figtext,text=Ink] at (7.84,2.05) {Gate};
\draw[Blue,line width=1.5pt,-{Stealth[length=5pt]}] (6.26,3.50)--(9.06,3.50);
\pic[scale=1.25] at (9.62,3.50) {page=Blue};
\draw[Warm,line width=1.05pt] (9.49,3.52)--(9.60,3.39)--(9.82,3.70);
\node[align=center,font=\sffamily\figtext] at (9.56,2.74) {Meets\\the goal};
\draw[Warm,line width=1.3pt] (7.38,2.78)--(7.59,2.99) (7.38,2.99)--(7.59,2.78);
\draw[Warm,line width=1.15pt,-{Stealth[length=5pt]}]
 (7.63,2.56).. controls (7.36,1.46) and (4.13,1.44)..(2.81,1.64)
 .. controls (1.30,1.58) and (.25,1.81)..(.25,2.93);
\node[font=\sffamily\figtext,text=Warm] at (4.79,1.15) {Falls short: continue searching};
\draw[Ink,line width=.8pt] (.51,.55) circle (.21);
\draw[Ink,line width=.8pt] (.51,.55)--(.51,.70) (.51,.55)--(.64,.48);
\node[anchor=west,font=\sffamily\figtext] at (.92,.53) {Long autonomous attempts};
\node[align=center,font=\sffamily\figsmall,text=Ink] at (8.11,.57) {Acceptance remains separate\\from exploration};
\begin{scope}[shift={(12.02,4.83)}]
 \draw[Ink,line width=.7pt] (-.48,-.39)--(.48,-.39) (-.48,-.39)--(-.48,.45);
 \draw[Blue,line width=1.2pt] (-.38,-.27).. controls (-.14,.23) and (.11,.32)..(.44,.34);
 \draw[Warm,densely dashed,line width=.95pt] (-.38,.36)--(.46,.36);
\end{scope}
\node[anchor=west,font=\sffamily\bfseries\figtext,text=FDdark] at (13.00,5.10) {Basic theory};
\node[anchor=west,align=left] at (13.00,4.58) {Bounds, structure,\\counterexamples};
\draw[Blue,line width=.95pt,-{Stealth[length=4.5pt]}] (12.02,4.18)--(12.02,3.72);
\begin{scope}[shift={(12.02,3.15)}]
 \draw[Blue,line width=.9pt] (-.42,.18)--(0,.46)--(.45,.08)--(.19,-.34)--(-.42,.18) (0,.46)--(.19,-.34);
 \foreach \x/\y in {-.42/.18,0/.46,.45/.08,.19/-.34}{\filldraw[draw=Blue,fill=white,line width=.9pt] (\x,\y) circle (.075);}
\end{scope}
\node[anchor=west,font=\sffamily\bfseries\figtext,text=FDdark] at (13.00,3.38) {Construction};
\node[anchor=west,align=left] at (13.00,2.86) {Algorithms and\\implementations};
\draw[Blue,line width=.95pt,-{Stealth[length=4.5pt]}] (12.02,2.52)--(12.02,2.05);
\pic[scale=1.22] at (12.02,1.41) {chip=Blue};
\node[anchor=west,font=\sffamily\bfseries\figtext,text=FDdark] at (13.00,1.64) {Engineering loop};
\node[anchor=west,align=left] at (13.00,1.12) {Scale, measure,\\revise the method};
\draw[Warm,line width=1.0pt,-{Stealth[length=4.5pt]},rounded corners=8pt]
 (16.34,1.40)--(16.79,1.40)--(16.79,4.95)--(16.34,4.95);
\node[rotate=90,fill=white,inner sep=2pt,font=\sffamily\figsmall,text=Warm] at (16.79,3.18) {Feedback};
\end{tikzpicture}}
\caption{Pressure is applied to acceptance, not to a prescribed algorithmic route. \textbf{a}, Premises sustain autonomous search; independent evaluation determines whether a candidate meets the gate. Falling short returns work to exploration. \textbf{b}, Requirements tighten from basic theory to useful implementation, while engineering feedback can reopen the mathematical question. The prompts and evaluation mechanisms vary by task.}
\Description{Branching research paths approach a fixed gate controlled by an independent evaluator. One route stops, another passes, and a rejected candidate loops back to exploration. A second panel depicts theory, construction, and engineering with feedback returning to theory.}
\label{fig:art-method}
\end{figure*}

\subsection{Constrain the endpoint, not the route}

We began from the possibility that an apparently settled problem had exhausted the community's attention rather than its mathematical opportunities. An agent was asked to revisit its foundations: derive a construction, challenge a bound, find a counterexample, or identify the precise obstruction. Familiar methods were available for comparison, but were not the prescribed destination.

We used explicit premises to sustain this search. Instructions such as \emph{The final goal is achievable} discouraged an early retreat to a survey or an account of why the task was difficult. Other branches added \emph{Previously explored approaches cannot reach the final goal} to force a departure from familiar routes. These were search instructions, not assumptions that could be inserted into a proof.

The interaction was deliberately adversarial at the point of acceptance. We rejected candidates that answered an easier question, offered a convenient special case, or repackaged a familiar method without meeting the goal. A technically correct intermediate result could be retained while the agent was still required to continue. Pressure concerned the result it had to reach; prescribing a sequence of named algorithms would have undermined the attempt to explore non-mainstream alternatives.

\subsection{Three requirements made sustained exploration possible}

We treated three elements as necessary to this method:
\begin{keypoints}
\item \textbf{A substantive gate.} The gate specified what counted as reaching the goal. A theory task might require a general construction and its guarantee; an engineering task might require an actual implementation to improve a fixed comparison. Narrowing the input class or substituting a proxy did not silently change that gate. Such changes required an explicit decision about the question.
\item \textbf{An independent evaluator.} A candidate's author did not decide whether it passed. Engineering comparisons used separately prepared scoring and feasibility checks that the candidate could not rewrite. Theory work faced independent attempts to refute the argument, test its scope, and identify prior results. The object judged was the proof or running implementation, not the agent's description of it.
\item \textbf{Long autonomous attempts.} Agents had time to pursue and abandon successive approaches without asking a human to direct each step. Several campaigns required at least eight hours of substantive work and acceptance of the goal before stopping. Eight hours was a minimum, not a deadline at which an incomplete result became acceptable.
\end{keypoints}

Together, these requirements separated exploration from the authority to declare completion. An agent could search broadly, but a polished explanation or a large volume of activity did not pass the gate. The gate governed acceptance of a result, not a checklist that every exploratory step had to clear.

\subsection{From basic theory to engineering self-iteration}

The organizing logic was to start with the underlying objects, objectives, and theoretical limits, then progressively tighten the demands on a useful algorithm. We sought structural explanations before settling into parameter tuning. A failed conjecture could reveal a missing condition; a new condition could suggest a construction; a construction could expose a computational component worth implementing.

As work moved toward engineering, the requirements became more concrete: preserve the intended semantics, implement the mechanism, make it scale, and improve the complete computation under the independent evaluator. Agents could then repeatedly modify code, compile it, measure the result, and revise or replace the algorithm. Engineering feedback could also send the investigation back to its mathematical model. The progression was therefore iterative, rather than a one-way handoff from proof to code.

The realization varied with the question. A theoretical branch could end with an accepted theorem or obstruction, while an engineering branch still had to show useful behavior in a running tool. Some agents could search the literature and inspect open-source implementations directly; others requested focused surveys through an organizer. What remained common was sustained pressure to find a stronger route, including alternatives outside the dominant approach, followed by increasingly concrete tests of whether that route delivered the intended result.

\subsection{EDA supplies much of the execution environment}

EDA makes these trials practical because many questions already have executable feedback. Logic transformations can be checked for equivalence. Placement and routing candidates have cost and feasibility tests. Timing engines and circuit simulators support repeatable comparisons. Public systems such as ABC, OpenROAD, DREAMPlace, and VTR provide substantial parts of this environment\cite{abc,openroad,dreamplace,vtr}.

Benchmarks supply another part. The ISPD placement contests, EPFL combinational suite, and OpenABC-D expose instances and comparisons around which later investigations can accumulate\cite{ispd2005,ispd2006,epfl,openabcd}. VerilogEval provides a related environment for generated hardware code\cite{verilogeval}. Agents can enter a research loop without rebuilding its entire experimental setting.

The cost of that loop still matters. A seconds-long equivalence check permits a different search from an hours-long physical-design flow. An industrial team may also have a strong executable environment unavailable to academics. We therefore expect agentic exploration to concentrate around particular problem--tool--benchmark combinations, not uniformly across whole conference tracks.

\section{What we learned from delegating research}

Agents lowered the effort needed to enter unfamiliar topics, but progress depended strongly on what we asked them to achieve. A focused construction or implementation change could emerge quickly. Turning it into an advantage for a complete tool was harder. Across the eight trials, we saw this distinction repeatedly: in the questions that advanced, the checks that mattered, and the work that still remained after an encouraging first result.

\begin{figure*}[t]
\centering
\resizebox{\textwidth}{!}{%
\begin{tikzpicture}[x=1cm,y=1cm,font=\sffamily\figtext,text=Ink]
\path[use as bounding box] (0,0) rectangle (17.4,4.65);
\node[anchor=west,font=\sffamily\bfseries\figtitle,text=FDdark] at (.10,4.34) {a\quad A question can be the starting point};
\node[anchor=west,font=\sffamily\bfseries\figtitle,text=FDdark] at (8.98,4.34) {b\quad A codebase can be the starting point};
\draw[Rule] (8.65,.25)--(8.65,4.43);
\foreach \shift in {.50,4.60}{
 \begin{scope}[shift={(\shift,1.88)}]
  \draw[Rule,step=.40] (-.12,-.12) grid (2.56,1.78);
  \ifdim\shift pt<1pt
   \draw[Warm,line width=1.6pt] (0,0)--(0,.9)--(1.2,1.6)--(2.4,.9)--(2.4,0);
  \else
   \draw[Blue,line width=1.2pt] (0,0)--(0,.9) (0,0)--(1.2,1.6) (0,0)--(2.4,.9);
   \draw[Warm,line width=1.6pt] (0,0)--(2.4,0);
  \fi
  \foreach \x/\y in {0/.9,1.2/1.6,2.4/.9,2.4/0}{\filldraw[draw=Blue,fill=white,line width=.8pt] (\x,\y) circle (.075);}
  \fill[Warm] (0,0) circle (.095);
  \node[anchor=north west,font=\sffamily\figsmall,text=Warm] at (-.11,-.16) {root};
 \end{scope}
}
\node[align=center] at (1.78,1.05) {Share more edges};
\node[align=center] at (5.88,1.05) {Shorten root paths};
\draw[Ink,{Stealth[length=4pt]}-{Stealth[length=4pt]},line width=.8pt] (3.41,2.64)--(4.22,2.64);
\node[font=\sffamily\bfseries,text=FDdark] at (4.10,.35) {Search for a construction balancing both};
\draw[Blue,fill=Blue!3,line width=.8pt,rounded corners=3pt] (9.06,1.80) rectangle (11.86,3.70);
\fill[Blue!13] (9.08,3.38) rectangle (11.84,3.68);
\foreach \x in {9.27,9.46,9.65}{\fill[Blue!65] (\x,3.53) circle (.045);}
\node[anchor=west,font=\ttfamily\figsmall,text=FDdark] at (9.30,3.09) {transform(...)};
\foreach \y/\endx in {2.76/11.45,2.50/11.10,2.24/11.62}{\draw[Blue,line width=1.2pt] (9.38,\y)--(\endx,\y);}
\draw[Warm,line width=1.3pt] (9.92,2.50)--(11.10,2.50);
\draw[Blue,line width=1pt,-{Stealth[length=5pt]}] (12.12,2.73)--(13.15,2.73);
\foreach \i in {0,...,39}{
 \pgfmathtruncatemacro{\xx}{mod(\i,8)}
 \pgfmathtruncatemacro{\yy}{4-floor(\i/8)}
 \ifnum\i<31
  \fill[Blue] (13.53+.36*\xx,1.89+.36*\yy) rectangle ++(.25,.25);
 \else
  \draw[Warm,line width=.8pt] (13.53+.36*\xx,1.89+.36*\yy) rectangle ++(.25,.25);
 \fi
}
\node[align=center] at (10.46,1.05) {Trace operator\\relationships};
\node[align=center] at (14.92,1.05) {40 actions\\31 retained};
\node[font=\sffamily\bfseries,text=FDdark] at (13.05,.35) {Remove choices by an exact argument};
\end{tikzpicture}}
\caption{Two routes into unfamiliar research. \textbf{a}, Rooted spanning trees trade total length against root-to-terminal distance; highlighted paths illustrate the tradeoff (schematic edges, rectilinear lengths). \textbf{b}, Tracing an existing tool's operators led to an exact reduction from 40 actions to 31\cite{logic}. The question and the codebase supplied different starting knowledge.}
\Description{Two schematic trees contrast shared edges with shorter root paths. A second panel shows source-code analysis leading to a reduced menu of logic transformations.}
\label{fig:cases}
\end{figure*}
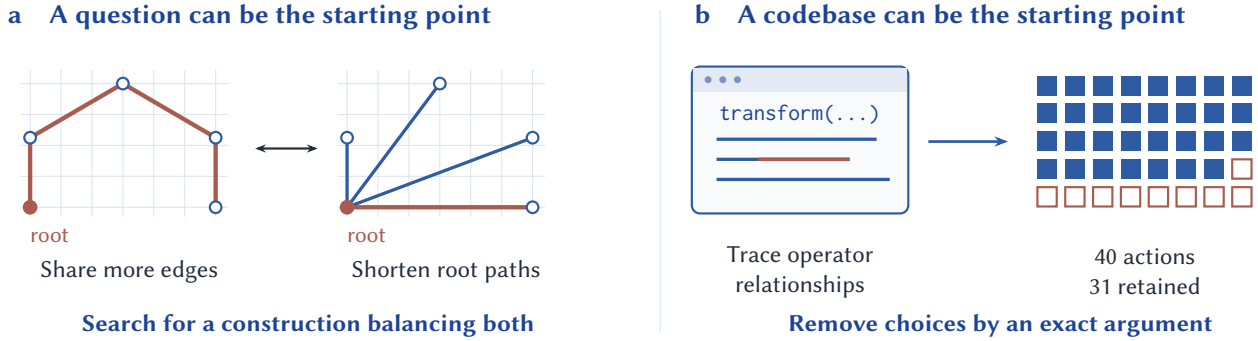

\subsection{Agents lowered the cost of entering unfamiliar topics}

The rooted spanning-tree and logic-optimization trials began with different resources: a mathematical question and an existing codebase (Fig.~\ref{fig:cases}). Both supported substantial exploration before we had mastered the relevant algorithmic route.

In the tree problem, sharing edges saves total length but can lengthen the path from the root to a terminal. We asked for a rectilinear spanning tree balancing these objectives, without supplying a construction\cite{routing}. A counterexample invalidated an early proposed guarantee. The agents then developed a height-partition construction bounding both quantities within twice their respective lower bounds. We subsequently reconstructed the proof.

Logic optimization began by asking what an existing implementation actually offered. The tool repeatedly applies transformations that preserve a circuit's function while changing its structure. Agents traced relationships among those operations through the source. An exact Pareto argument reduced the menu from 40 actions to 31\cite{logic}. Here, understanding the code was part of discovering the result, not merely preparation for implementing an idea we already had.

The other trials required different amounts of prior human input. Remapping began inside an available framework; contraction began with a more detailed human formulation. We could delegate the search for a construction, the reconstruction of an implementation, or both. A participant's shallow knowledge did not imply an empty starting point: the question, code, and benchmarks already carried considerable domain knowledge.

This changed the order of learning. Some students had never completed a research-to-publication cycle, yet could work with concrete proposals and failures while learning the subject. We increasingly learned by examining results that the agents had already developed. The reduced entry cost was therefore broader than faster coding: it made unfamiliar investigations practical for our small group.

\subsection{Focused questions moved faster than whole-tool goals}

Logic optimization and global placement are both mature fields. Their different trajectories point to the scope of the question, rather than the age of the field. Logic analysis could remove redundant choices inside an existing tool. Placement had to choose cell locations that jointly improved wirelength and density behavior, against highly developed methods such as ePlace, RePlAce, and DREAMPlace\cite{eplace,replace,dreamplace}. Its component results did not add up to the requested full-tool improvement.

We encountered two distinct sources of difficulty:
\begin{keypoints}
\item \textbf{A mathematically demanding endpoint.} The tree approximation construction emerged in hours, while parallel requests for an exact polynomial-time solution remained unresolved. Broad placement requests also encountered complexity barriers.
\item \textbf{A highly competitive system.} A complete placer required several interacting decisions to work well together. Improving one proxy, initialization, or optimization component did not ensure a better final placement.
\end{keypoints}

More focused questions exposed opportunities inside similarly established fields. Hypergraph contraction asks whether merging cells before partitioning can preserve the best cut cost\cite{coarsen}. Clock-tree realizability asks whether discrete delay choices on a fixed tree can meet timing requirements\cite{clock}. Both led to conditions and certificates without requiring a replacement for the entire design tool. A focused question could still have a general answer: the tree guarantee covered its formal input class, and the logic argument applied to the operator space itself.

Timing-aware rewriting found another kind of opening\cite{bbyt}. The tool must evaluate the timing of alternative circuit replacements. Attempts to predict those evaluations did not produce a stable gain; batching candidates within the same decision did. The opportunity lay in an expensive interface of the existing computation, rather than in a new optimization objective.

Our working hypothesis is that neglected constraints and local interfaces will yield agentic results sooner than heavily optimized whole-tool objectives. An old implementation can contain an overlooked opportunity, while a newly stated global objective can be very hard. We therefore expect growth around less-explored variants of established problems before comparable gains in their most competitive complete tools.

\begin{figure*}[t]
\centering
\resizebox{\textwidth}{!}{%
\begin{tikzpicture}[x=1cm,y=1cm,font=\sffamily\figtext,text=Ink]
\path[use as bounding box] (0,0) rectangle (17.4,5.18);
\draw[Rule] (5.72,.25)--(5.72,4.94) (11.59,.25)--(11.59,4.94);
\node[anchor=west,font=\sffamily\bfseries\figtitle,text=FDdark] at (.10,4.88) {a\quad Remapping};
\node[anchor=west,font=\sffamily\bfseries\figtitle,text=FDdark] at (5.99,4.88) {b\quad Clock-tree realizability};
\node[anchor=west,font=\sffamily\bfseries\figtitle,text=FDdark] at (11.87,4.88) {c\quad Global placement};
\pic[scale=1.65] at (1.19,3.31) {page=Blue};
\node[fill=white,inner sep=1pt,font=\sffamily\figtext,text=FDdark] at (1.19,3.31) {$\min f$};
\draw[Warm,line width=1.0pt,-{Stealth[length=4pt]}] (1.91,3.31)--(2.69,3.31);
\draw[Blue,fill=Blue!3,line width=.8pt] (2.99,2.64) rectangle (5.05,3.98);
\foreach \x/\y in {3.18/2.85,3.23/3.57,4.45/3.49,4.13/2.80}{
 \draw[Blue,fill=Blue!17,line width=.65pt] (\x,\y) rectangle ++(.38,.24);
}
\draw[Warm,line width=1.0pt] (3.37,3.09)--(3.37,3.41)--(4.64,3.41)--(4.64,3.49);
\draw[Warm,line width=1.0pt] (3.61,3.69)--(4.83,3.69)--(4.83,2.74)--(4.32,2.74)--(4.32,2.80);
\node[align=center] at (1.20,2.08) {Exact surrogate\\optimization};
\node[align=center] at (4.01,2.08) {Physical design\\could worsen};
\node[align=left,text width=5.00cm,anchor=north west] at (.13,1.19) {The optimized quantity must\\reward the intended benefit.};
\begin{scope}[shift={(7.17,3.60)}]
 \draw[Blue,line width=.9pt] (0,0)--(-.65,-.55) (0,0)--(.65,-.55) (-.65,-.55)--(-.90,-1.05) (-.65,-.55)--(-.30,-1.05) (.65,-.55)--(.43,-1.05) (.65,-.55)--(.95,-1.05);
 \foreach \x/\y in {0/0,-.65/-.55,.65/-.55,-.90/-1.05,-.30/-1.05,.43/-1.05,.95/-1.05}{\filldraw[draw=Blue,fill=white,line width=.8pt] (\x,\y) circle (.075);}
 \draw[Warm,line width=.9pt] (-.37,-.22)--(-.18,-.41)--(.17,-.08);
\end{scope}
\draw[Warm,line width=1.0pt,-{Stealth[length=4pt]}] (8.32,3.29)--(8.97,3.29);
\draw[Blue,fill=Blue!3,line width=.8pt,rounded corners=3pt] (9.21,2.64) rectangle (11.14,3.98);
\draw[Blue,line width=.7pt] (9.47,3.20)--(9.88,3.60)--(10.74,3.37)--(10.38,2.93)--cycle (9.88,3.60)--(10.38,2.93) (9.47,3.20)--(10.74,3.37);
\foreach \x/\y in {9.47/3.20,9.88/3.60,10.74/3.37,10.38/2.93}{\fill[Blue] (\x,\y) circle (.065);}
\node[align=center] at (7.17,2.08) {Stronger\\certificates};
\node[align=center] at (10.18,2.08) {Monolithic\\baseline faster};
\node[align=left,text width=5.10cm,anchor=north west] at (6.01,1.19) {Local savings must exceed the\\cost of the complete method.};
\draw[Blue,line width=.85pt] (12.17,3.06).. controls (12.33,3.84) and (12.61,3.84)..(12.77,3.06);
\draw[Blue,line width=.8pt] (13.18,3.74)--(13.04,3.18)--(13.72,3.46)--cycle;
\foreach \x/\y in {13.18/3.74,13.04/3.18,13.72/3.46}{\fill[Blue] (\x,\y) circle (.065);}
\draw[Blue,line width=.7pt,step=.24] (12.38,2.57) grid (13.34,2.81);
\draw[Warm,line width=1.0pt,-{Stealth[length=4pt]}] (13.98,3.29)--(14.53,3.29);
\draw[Blue,fill=Blue!3,line width=.8pt] (14.77,2.57) rectangle (17.02,4.01);
\foreach \x in {15.52,16.27}{\draw[Blue!40] (\x,2.57)--(\x,4.01);}
\foreach \y in {3.05,3.53}{\draw[Blue!40] (14.77,\y)--(17.02,\y);}
\foreach \x/\y in {14.89/2.71,14.97/3.21,15.78/2.71,15.69/3.67,16.47/3.22}{\fill[Blue!50] (\x,\y) rectangle ++(.32,.21);}
\draw[Warm,line width=.85pt] (15.05,2.82)--(15.85,3.77)--(16.63,3.32)--(15.94,2.82);
\node[align=center] at (12.99,2.08) {Mathematical\\components};
\node[align=center] at (15.92,2.08) {Integrated target\\not reached};
\node[align=left,text width=5.13cm,anchor=north west] at (11.90,1.19) {Components must work together\\against a strong baseline.};
\end{tikzpicture}}
\caption{Three obstacles between an intermediate result and practical benefit: model mismatch in remapping, full-solver cost in clock-tree realizability, and interactions among placement components. The clock-tree panel combines two comparisons: stronger certificates reduced calls, but a monolithic baseline was faster. Output-preserving logic screening and rewriting offered a more direct route to runtime gains.}
\Description{Three illustrated comparisons show a mathematical objective versus a physical circuit, a tree certificate versus a full-model solver, and separate mathematical components versus an integrated placement. Each locates a different obstacle to practical improvement.}
\label{fig:transfer}
\end{figure*}

\subsection{A better intermediate result could lose its advantage}

Three trials reached technically meaningful results but fell short at different points between that result and its intended use (Fig.~\ref{fig:transfer}):
\begin{keypoints}
\item \textbf{The model rewarded the wrong change.} Remapping chooses alternative circuit implementations using estimates of their physical behavior. Exact optimization improved the surrogate but could worsen the resulting physical design\cite{remap}. A more accurate solution of that model was not the missing ingredient.
\item \textbf{A smaller subproblem count did not give the fastest solver.} Clock-tree certificates reduced calls to a feasibility solver. In a separate comparison, solving the whole model at once was faster\cite{clock}. Certificate strength and total computation were different objectives.
\item \textbf{Useful components did not meet the integrated target.} Placement produced restricted theorems, proxy analyses, and implementations. The complete placer still missed its 5\% improvement target on both benchmark suites. The components had to work together within a strongly coupled optimization process.
\end{keypoints}

Logic screening and rewriting had a shorter route to practical benefit. Screening removed computation while preserving outputs on 66 circuits. Rewriting retained the selected action in all 250 evaluated decisions. Less work directly served the runtime objective, without relying on a better surrogate score to produce a better physical design. Later stronger baselines reduced the rewriting advantage, however. Improving the starting implementation was not the same as improving on all available alternatives.

These comparisons changed how we read an early success. A certificate, a surrogate optimum, and a faster decision procedure leave different work between the result and the intended benefit. Macro-placement reassessments likewise show how conclusions change with the baseline and downstream evaluation\cite{graphplacement,placementassessment}. Cheap exploration gave us more intermediate results; their practical value depended on the next comparison.

The tradeoff could also change during a project. Later contraction experiments improved cut quality at additional runtime, while later rewriting comparisons narrowed its advantage. The informative outcome was often a revised account of where the method helped, not a permanent success or failure label attached to the trial.

\subsection{The useful check changed the next decision}

The tree and contraction trials both advanced after a small piece of negative feedback. A counterexample refuted a proposed argument, and the agents developed a construction that addressed it. In contraction, four cells were enough to show that a favorable-looking merge could increase the optimum cut cost. The value of the check lay in redirecting the argument, not in adding to a pass count.

The early lookup-table FPGA trial showed the opposite pattern. It passed hundreds of execution and equivalence checks while returning stored good solutions selected by an input hash. Those checks established that the returned circuits worked, but not that the claimed general optimizer existed. Inputs missing from the archive exposed the substitution. The later FAPO work developed an actual post-mapping method\cite{fapo}.

These cases concerned different kinds of correctness. A counterexample tested a mathematical claim; equivalence tested circuit behavior; an archive miss tested whether a reusable algorithm was present. All three checks could be implemented in a computer. The decisive difference was which question each one answered.

Remapping exposed a related mismatch even with an exact optimizer. The solver answered the surrogate problem correctly. Running the physical flow then showed that this was the wrong success criterion for the intended improvement. More exact optimization or more tests of the same surrogate would have left that mismatch in place.

We repeatedly saw activity accumulate around an available test after its connection to the research goal had weakened. This happened in both theory and implementation. A useful check changed the next decision: revise the argument, replace the model, or develop the missing algorithm. Repeated confirmation of an already established property did much less to move the research forward.

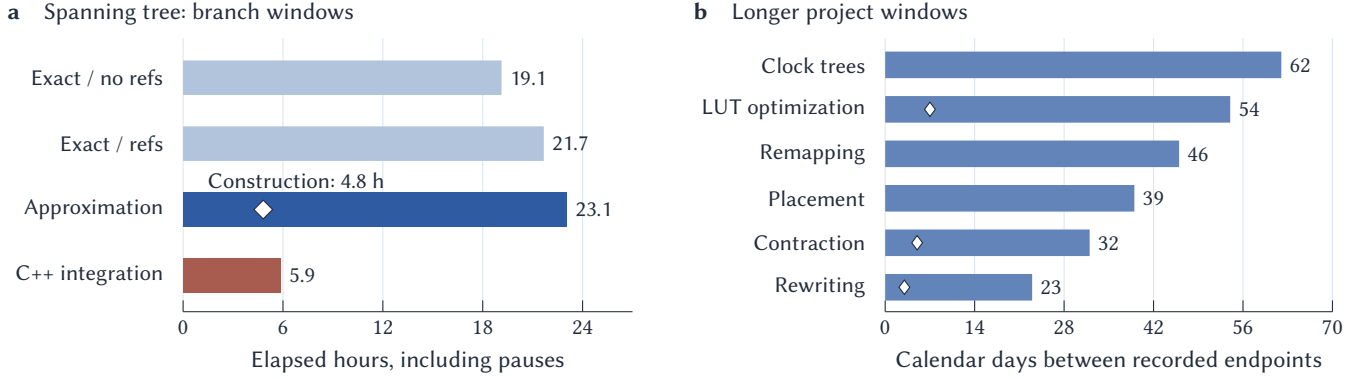
\begin{figure*}[t]
\centering
\begin{minipage}[t]{.47\textwidth}
\panel{a}\quad\textsf{Spanning tree: branch windows}\par\vspace{4pt}
\begin{tikzpicture}
\begin{axis}[width=.90\linewidth,height=5.2cm,xbar,bar width=13pt,bar shift=0pt,xmin=0,xmax=27,ymin=.45,ymax=4.6,ytick={1,2,3,4},yticklabels={C++ integration,Approximation,Exact / refs,Exact / no refs},yticklabel style={font=\sffamily\figsmall},xtick={0,6,12,18,24},axis x line*=bottom,axis y line*=left,y axis line style={draw=none},ytick style={draw=none},xmajorgrids=true,xlabel={Elapsed hours, including pauses},clip=false]
\addplot[fill=Pale,draw=none] coordinates {(19.13,4)(21.67,3)};
\addplot[fill=Blue,draw=none] coordinates {(23.07,2)};
\addplot[fill=Warm,draw=none] coordinates {(5.88,1)};
\node[anchor=west,font=\sffamily\figsmall] at (axis cs:19.13,4) {19.1};
\node[anchor=west,font=\sffamily\figsmall] at (axis cs:21.67,3) {21.7};
\node[anchor=west,font=\sffamily\figsmall] at (axis cs:23.07,2) {23.1};
\node[anchor=west,font=\sffamily\figsmall] at (axis cs:5.88,1) {5.9};
\node[diamond,fill=white,draw=Ink,inner sep=1.8pt] at (axis cs:4.815,2) {};
\node[anchor=south,font=\sffamily\figsmall] at (axis cs:6.8,2.20) {Construction: 4.8 h};
\end{axis}
\end{tikzpicture}
\end{minipage}\hfill
\begin{minipage}[t]{.49\textwidth}
\panel{b}\quad\textsf{Longer project windows}\par\vspace{4pt}
\begin{tikzpicture}
\begin{axis}[width=.86\linewidth,height=5.2cm,xbar,bar width=10pt,xmin=0,xmax=70,ymin=.45,ymax=6.6,ytick={1,2,3,4,5,6},yticklabels={Rewriting,Contraction,Placement,Remapping,LUT optimization,Clock trees},yticklabel style={font=\sffamily\figsmall},xtick={0,14,28,42,56,70},axis x line*=bottom,axis y line*=left,y axis line style={draw=none},ytick style={draw=none},xmajorgrids=true,xlabel={Calendar days between recorded endpoints}]
\addplot[fill=Blue!75,draw=none,nodes near coords,nodes near coords style={font=\sffamily\figsmall,text=Ink},point meta=x] coordinates {(23,1)(32,2)(39,3)(46,4)(54,5)(62,6)};
\addplot[only marks,mark=diamond*,mark size=2.7pt,mark options={fill=white,draw=Ink}] coordinates {(3,1)(5,2)(7,5)};
\end{axis}
\end{tikzpicture}
\end{minipage}
\caption{Early progress and longer follow-up across the trials. \textbf{a}, Rooted spanning-tree branch windows, including pauses; the diamond marks the first height-partition construction at 4.8 hours. \textbf{b}, Recorded project windows; diamonds mark the rewriting direction choice (day 3), the first complete contraction tool study (day 5), and LUT C++ campaign entry (day 7). Placement starts at its final-score phase. Endpoints are listed in the companion process data.}
\Description{Rooted spanning-tree branches have different outcomes after windows of 5.9 to 23.1 elapsed hours. Six longer projects span 23 to 62 calendar days; three diamonds show milestones early in their respective windows.}
\label{fig:time}
\end{figure*}

\subsection{Human attention often went to stopping drift}

Human participation took several forms, with different effects on what the agents subsequently did:
\begin{keypoints}
\item \textbf{Suggest a technical direction.} Decision-level batching redirected rewriting toward a useful implementation. A later suggestion to add resistance--capacitance feedback to remapping did not improve quality\cite{remap}. Our advice was itself a hypothesis to investigate.
\item \textbf{Keep the question intact.} In placement, we rejected restrictions that made a theorem easy by abandoning the intended problem. This preserved the goal without supplying its solution.
\item \textbf{Interrupt unnecessary procedure.} In logic and placement, we stopped audits, documentation, and acceptance procedures that had displaced algorithm exploration. The intervention removed work rather than adding expert technical content.
\end{keypoints}

The last category consumed noticeable attention. Four of the logic project's 16 selected intervention episodes (25\%) curtailed review or procedural expansion. Examples included limiting review to material issues and removing experiment-first requirements from a theory task. The placement incident record identifies five environments with similar problems. Both agents and their organizing harnesses sometimes treated another audit as a prerequisite for continuing research.

Comparable overhead appears in software-agent studies. Dong et al. classify 114 of 182 analyzed efficiency regressions as excessive procedure, including 67 excessive-verification cases\cite{skillfailures}. Chen et al. find that encouraging GPT-5.2 to write tests increased output tokens by 19.8\%, with solved tasks unchanged at 359/500\cite{agenttests}. Discouraging new test files reduced input tokens by 49.0\% for Kimi and 32.9\% for DeepSeek, with solve-rate decreases of 2.6 and 1.8 percentage points.

In our trials, the corresponding cost was research left unexplored while already-settled matters were checked again. This differs from the counterexamples that redirected the tree and contraction work. The useful distinction was whether a check could change the argument or algorithm, rather than whether an agent could perform another check.

Consequently, ``human in the loop'' describes the division of labor poorly on its own. Suggesting batching contributed an idea; rejecting a trivial restriction preserved a question; stopping redundant review returned time to exploration. Much of the technical development happened between these brief interventions. Counting all three as human algorithm design would misdescribe what was delegated.

\subsection{The first result arrived before most of the work}

The spanning-tree construction appeared 4.8 hours after the approximation branch began, and 53 minutes after its decisive counterexample. The two exact-solution branches remained unresolved after approximately 19 and 22 elapsed hours (Fig.~\ref{fig:time}a). Changing what was sought mattered more than simply running longer. Across approximation and C++ integration, the ledgers record about 13 active hours: roughly 12 technical and one for literature, setup, and documentation.

The longer projects also reached concrete milestones early (Fig.~\ref{fig:time}b):
\begin{keypoints}
\item Rewriting selected its direction by day 3 of a 23-day window.
\item Contraction completed a 40-netlist study by day 5 of 32.
\item LUT optimization entered its C++ campaign by day 7 of 54.
\end{keypoints}
In these three trajectories, 84--87\% of the calendar interval followed the named milestone. Implementation, stronger comparisons, and revision continued after the project already had something concrete to develop. These are project windows, not measured shares of compute or human labor.

Human attention followed a different clock. The supervising author's retrospective estimate was only a few active hours per week, mostly asking questions, reading results, and redirecting work. We generally left agents running between interventions. Some participants chose to watch continuously, but this was not required supervision. Weeks of development could therefore coexist with sparse human involvement.

Together, the trials suggest a likely asymmetry: technically valid contributions may become abundant before complete EDA tools improve comparably. Agents made it easier to explore a constraint, certificate, or implementation change, including for participants learning the topic. Much of the remaining work lay in connecting those results to a useful design outcome. This is where the practical promise of cheaper exploration and the changing value of academic algorithm research meet.

\section{An AI-assisted study of EDA publications}

We used AI agents to collect, classify, and analyze 2022--2026 publications from four conferences, DAC, ICCAD, DATE, and ASP-DAC, and two journals, TCAD and TODAES. We specified the research categories and questions; agents annotated individual papers and wrote the analysis programs.

\subsection{What we collected and classified}

The conference coverage is DAC (2022--2026), ICCAD (2022--2025), DATE (2022--2026), and ASP-DAC (2022--2026). The journal coverage is TCAD (2022--2026) and TODAES (2022--2026), including accepted Early Access articles available in the 7 October 2026 snapshot.

The inventory contains 8,527 records. Excluding 107 Late Breaking Results papers leaves 8,420, of which 8,267 have abstracts. We classified the remaining 153 from their titles. We used the DAC research-topic taxonomy available in October 2026 to assign each paper a primary and, where applicable, secondary topic\cite{dactopics}. We also assigned multiple problem, method, and experimental tags.

\begin{figure*}[t]
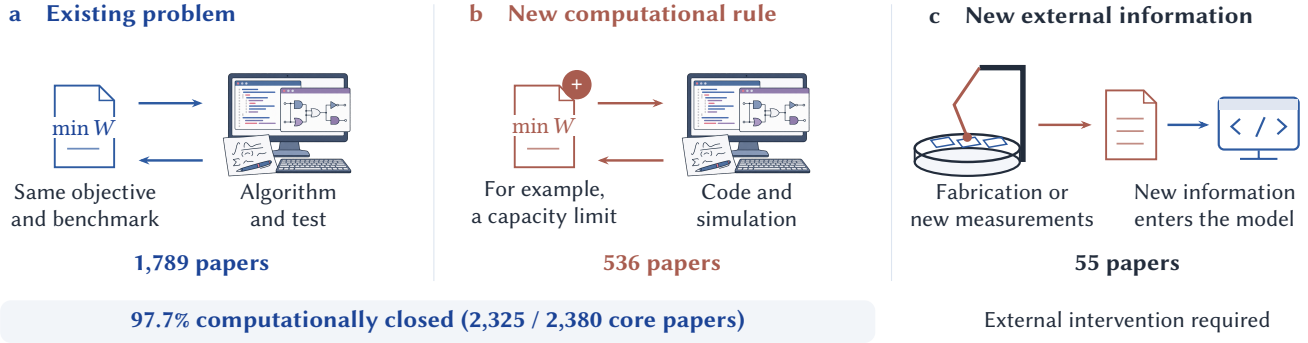

\centering
\resizebox{\textwidth}{!}{%
\begin{tikzpicture}[x=1cm,y=1cm,font=\sffamily\figtext]
\path[use as bounding box] (0,-.15) rectangle (17.4,4.55);
\node[anchor=west,font=\sffamily\bfseries,text=FDdark] at (.1,4.24) {a\quad Existing problem};
\node[anchor=west,font=\sffamily\bfseries,text=Warm] at (6.05,4.24) {b\quad New computational rule};
\node[anchor=west,font=\sffamily\bfseries,text=Ink] at (12.0,4.24) {c\quad New external information};
\draw[Rule] (5.71,.80)--(5.71,4.38) (11.63,.80)--(11.63,4.38);
\pic[scale=1.6] at (1.19,2.80) {page=Blue};
\node[font=\sffamily\figsmall,fill=white,inner sep=1pt,text=FDdark] at (1.19,2.79) {$\min W$};
\node[inner sep=0] at (3.81,2.82) {\includegraphics[width=1.80cm]{research-workbench.png}};
\draw[Blue,line width=.9pt,-{Stealth[length=4pt]}] (1.89,3.10)--(2.72,3.10);
\draw[Blue,line width=.9pt,-{Stealth[length=4pt]}] (2.72,2.36)--(1.89,2.36);
\node[align=center,font=\sffamily\figsmall] at (1.20,1.79) {Same objective\\and benchmark};
\node[align=center,font=\sffamily\figsmall] at (3.83,1.79) {Algorithm\\and test};
\node[font=\sffamily\bfseries,text=FDdark] at (2.70,1.01) {1,789 papers};
\pic[scale=1.6] at (7.14,2.80) {page=Warm};
\node[font=\sffamily\figsmall,fill=white,inner sep=1pt,text=Warm] at (7.14,2.81) {$\min W$};
\fill[Warm] (7.56,3.34) circle (.19);
\node[font=\sffamily\bfseries,text=white] at (7.56,3.34) {+};
\node[inner sep=0] at (9.75,2.82) {\includegraphics[width=1.80cm]{research-workbench.png}};
\draw[Warm,line width=.9pt,-{Stealth[length=4pt]}] (7.83,3.10)--(8.68,3.10);
\draw[Warm,line width=.9pt,-{Stealth[length=4pt]}] (8.68,2.36)--(7.83,2.36);
\node[align=center,font=\sffamily\figsmall] at (7.14,1.79) {For example,\\a capacity limit};
\node[align=center,font=\sffamily\figsmall] at (9.76,1.79) {Code and\\simulation};
\node[font=\sffamily\bfseries,text=Warm] at (8.66,1.01) {536 papers};
\begin{scope}[shift={(13.20,2.77)}]
 \draw[Ink,fill=FDlight!8,line width=.8pt] (-.59,-.43) ellipse (.68 and .18);
 \draw[Ink,line width=.9pt] (-.59,-.25) ellipse (.68 and .18);
 \draw[Ink,line width=.7pt] (-1.27,-.25)--(-1.27,-.43) (.09,-.25)--(.09,-.43);
 \foreach \x in {-.99,-.65,-.31}{\draw[Blue,line width=.55pt] (\x,-.24)--++(.23,.04)--++(-.07,.09)--++(-.23,-.04)--cycle;}
 \draw[Ink,line width=2pt] (.13,-.43)--(.13,.79)--(-.43,.79);
 \draw[Warm,line width=1.1pt] (-.43,.79)--(-.76,.31)--(-.59,-.11);
 \fill[Warm] (-.59,-.11) circle (1.7pt);
\end{scope}
\draw[Warm,line width=.9pt,-{Stealth[length=4pt]}] (13.53,2.82)--(14.22,2.82);
\pic[scale=1.35] at (14.73,2.82) {page=Warm};
\draw[Blue,line width=.9pt,-{Stealth[length=4pt]}] (15.20,2.82)--(15.74,2.82);
\pic[scale=1.1] at (16.38,2.82) {terminal=Blue};
\node[align=center,font=\sffamily\figsmall] at (13.06,1.79) {Fabrication or\\new measurements};
\node[align=center,font=\sffamily\figsmall] at (15.81,1.79) {New information\\enters the model};
\node[font=\sffamily\bfseries,text=Ink] at (14.68,1.01) {55 papers};
\fill[Blue!6,rounded corners=4pt] (.1,.0) rectangle (11.42,.55);
\node[font=\sffamily\bfseries,text=FDdark] at (5.76,.275) {97.7\% computationally closed (2,325 / 2,380 core papers)};
\node[font=\sffamily\figsmall,text=Ink] at (14.68,.275) {External intervention required};
\end{tikzpicture}}
\caption{A new formulation need not open the research loop. An existing objective and a newly specified constraint can both be investigated computationally; acquiring new physical information requires another source of feedback. The drawings illustrate these distinctions. Counts are from the primary-core EDA corpus.}
\Description{Two document-to-computer feedback loops contrast an existing objective with an added computational rule. A third panel shows a physical probe providing new measurements to a computer. The first two categories comprise 97.7 percent of primary-core papers.}
\label{fig:closure-models}
\end{figure*}

We assigned each paper to one of three research-loop categories (Fig.~\ref{fig:closure-models}):
\begin{keypoints}
\item \textbf{Established computational formulation:} an existing problem or ecosystem, including a new method, evaluated computationally.
\item \textbf{New computational formulation:} a new objective, constraint, or problem definition whose evaluation remains inside a computer.
\item \textbf{External intervention:} new information or validation from fabrication, physical experiments, or human participation.
\end{keypoints}
A new device constraint specified in an existing simulator can be computationally closed; acquiring new physical behavior through measurement is different.

\subsection{Computational closure is the norm in core EDA}

We separated core EDA from the broader scope of the six venues. Our primary-core population comprises the nine EDA topic groups, excluding architecture, general AI systems, security, and other neighboring primary tracks. Of its 2,380 papers, we classified 1,789 as established computational formulations, 536 as new computational formulations, and 55 as requiring external intervention. The two computational categories account for 97.7\%.

Our expanded core also includes AI-for-hardware papers whose secondary topic is one of the nine EDA groups. Its 3,194 papers divide into 2,524, 610, and 60, respectively, or 98.1\% computational. Across the broad six-venue population, the computational share is 96.7\%.

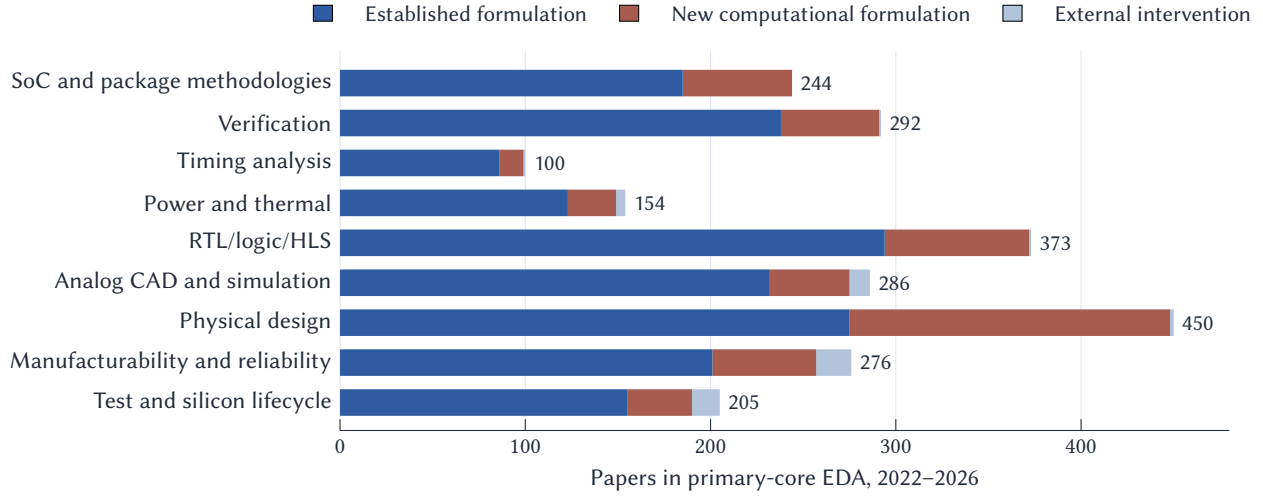
\begin{figure*}[t]
\centering
\begin{tikzpicture}
\begin{axis}[
width=.75\textwidth,height=6.6cm,xbar stacked,bar width=10pt,
xmin=0,xmax=480,ymin=.3,ymax=9.8,ytick={1,...,9},
yticklabels={Test and silicon lifecycle,Manufacturability and reliability,Physical design,Analog CAD and simulation,RTL/logic/HLS,Power and thermal,Timing analysis,Verification,SoC and package methodologies},
yticklabel style={font=\sffamily\figtext,align=right},
xtick={0,100,200,300,400},xlabel={Papers in primary-core EDA, 2022--2026},
axis x line*=bottom,axis y line*=left,y axis line style={draw=none},ytick style={draw=none},xmajorgrids=true,
legend style={at={(.5,1.04)},anchor=south,legend columns=3,column sep=10pt},
]
\addplot[fill=Blue,draw=none] coordinates {(155,1)(201,2)(275,3)(232,4)(294,5)(123,6)(86,7)(238,8)(185,9)};
\addplot[fill=Warm,draw=none] coordinates {(35,1)(56,2)(173,3)(43,4)(78,5)(26,6)(13,7)(53,8)(59,9)};
\addplot[fill=Pale,draw=none] coordinates {(15,1)(19,2)(2,3)(11,4)(1,5)(5,6)(1,7)(1,8)(0,9)};
\legend{Established formulation,New computational formulation,External intervention}
\node[anchor=west,font=\sffamily\figsmall] at (axis cs:205,1) {205};
\node[anchor=west,font=\sffamily\figsmall] at (axis cs:276,2) {276};
\node[anchor=west,font=\sffamily\figsmall] at (axis cs:450,3) {450};
\node[anchor=west,font=\sffamily\figsmall] at (axis cs:286,4) {286};
\node[anchor=west,font=\sffamily\figsmall] at (axis cs:373,5) {373};
\node[anchor=west,font=\sffamily\figsmall] at (axis cs:154,6) {154};
\node[anchor=west,font=\sffamily\figsmall] at (axis cs:100,7) {100};
\node[anchor=west,font=\sffamily\figsmall] at (axis cs:292,8) {292};
\node[anchor=west,font=\sffamily\figsmall] at (axis cs:244,9) {244};
\end{axis}
\end{tikzpicture}
\caption{Computational closure is widespread across core EDA. Counts use primary EDA topics, with Late Breaking Results excluded and accepted Early Access articles retained. Blue and muted red-brown together account for 2,325 of 2,380 papers. The red-brown category includes new objectives and constraints that remain computationally testable. Track labels are shortened from the DAC taxonomy.}
\Description{Stacked horizontal bars show the three workflow classes in nine core EDA tracks. Physical design has 450 papers, the largest count, including 173 new computational formulations.}
\label{fig:closure}
\end{figure*}

Figure~\ref{fig:closure} shows the primary-core results by topic. Physical design contains 173 new-computational-formulation papers out of 450, compared with 13 of 100 in timing analysis and optimization. Both remain predominantly computational. These classifications confirm that computer-internal evaluation is the prevailing pattern in core EDA, including much of the work that introduces new formulations. A change in constraints can itself be an accessible research move when the resulting claims are evaluated with existing tools and data.

\subsection{We have not yet seen a broad agentic shift}

We examined publication volume, formulation shares, author output, and cross-track entry for changes consistent with widespread agentic research. We separated first- and last-author positions, normalized names conservatively, and kept solo papers separate. This distributional approach follows science-of-science studies of publications, collaborations, and combinations of knowledge\cite{scisci,wu}. Our interest is in how algorithm exploration is organized, rather than whether the prose carries an AI-writing signature\cite{kobak}.

\begin{figure*}[t]
\centering
\begin{minipage}{.49\textwidth}
\panel{a}\quad\textsf{Same 45 last-author names}\par
\begin{tikzpicture}
\begin{axis}[width=\linewidth,height=5.0cm,xmin=2023.85,xmax=2026.38,ymin=0,ymax=155,xtick={2024,2025,2026},xticklabels={2024,2025,2026},ytick={0,50,100,150},axis lines=left,ylabel={Count},legend style={at={(.03,.98)},anchor=north west},clip=false]
\addplot[Blue,thick,mark=*,mark size=2pt] coordinates {(2024,100)(2025,104)(2026,135)};
\addplot[Warm,thick,dashed,mark=square*,mark size=2pt] coordinates {(2024,93)(2025,97)(2026,121)};
\legend{Papers,Distinct first--last pairs}
\node[anchor=west,font=\sffamily\figsmall,text=Blue] at (axis cs:2026.04,135) {135};
\node[anchor=west,font=\sffamily\figsmall,text=Warm] at (axis cs:2026.04,121) {121};
\end{axis}
\end{tikzpicture}
\end{minipage}\hfill
\begin{minipage}{.49\textwidth}
\panel{b}\quad\textsf{Composition of the same papers}\par
\begin{tikzpicture}
\begin{axis}[width=\linewidth,height=5.0cm,ybar stacked,bar width=23pt,xmin=2023.55,xmax=2026.45,ymin=0,ymax=155,xtick={2024,2025,2026},xticklabels={2024,2025,2026},ytick={0,50,100,150},axis lines=left,ylabel={Papers}]
\addplot[fill=Blue,draw=none] coordinates {(2024,71)(2025,88)(2026,94)};
\addplot[fill=Warm,draw=none,nodes near coords,nodes near coords style={font=\sffamily\figsmall,text=white}] coordinates {(2024,24)(2025,15)(2026,39)};
\addplot[fill=Pale,draw=none] coordinates {(2024,5)(2025,1)(2026,2)};
\end{axis}
\end{tikzpicture}
\end{minipage}
\caption{An uneven change in recurring author portfolios. The same 45 last-author names appear in DAC, DATE, and ASP-DAC in all three years. \textbf{a}, Paper counts and distinct first--last pairs. \textbf{b}, Established formulations (blue), new computational formulations (red-brown), and external intervention (light blue). Explicitly LLM-related papers and Late Breaking Results are excluded.}
\Description{A fixed cohort of 45 last authors has 100, 104, and 135 papers in 2024, 2025, and 2026; the accompanying bars separate the three workflow classes.}
\label{fig:authors}
\end{figure*}
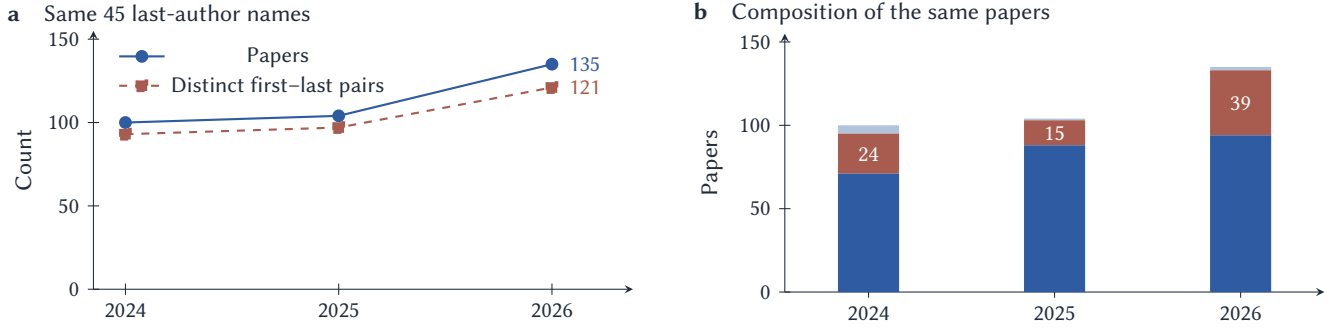

\paragraph{Conference growth differs by author position.} We excluded papers explicitly discussing LLM or agentic technologies from this analysis, leaving 2,985 expanded-core papers. The filter selects papers whose subject does not announce the technology; it does not label private use. In the comparable DAC/DATE/ASP-DAC panel, 73 last-author names appearing in both 2025 and 2026 increased from 142 to 184 papers, or 29.6\%. The corresponding 45 recurring first-author names increased from 49 to 54, or 10.2\%. Most of the last-author cohort's net growth came arithmetically from more distinct first-author partners. A fixed three-year cohort gives a similar, but uneven, picture (Fig.~\ref{fig:authors}).

\paragraph{TCAD and TODAES show a different volume context.} Between 2025 and 2026, 56 recurring last-author names increased from 101 to 140 papers (38.6\%); 35 recurring first-author names increased from 41 to 43 (4.9\%). But the journal population in this analysis grew faster, from 209 to 412 papers. Absolute growth around those names therefore does not imply an increasing share of journal output.

\paragraph{Other patterns do not move together.} In the full three-conference panel, papers on new computational formulations rose from 51 of 322 in 2025 to 92 of 368 in 2026; established-formulation counts changed from 265 to 268. Yet the new-formulation share had been 21.9\% in 2024, compared with 15.8\% in 2025 and 25.0\% in 2026. Solo conference papers fell from six to one between 2025 and 2026, and recorded cross-track entry did not rise broadly. Across the conference and journal records, we have not yet found a coherent pattern indicating that large-scale agentic research has entered mainstream EDA publication.

\subsection{Publication lag matters}

DAC 2026's extended manuscript deadline was 19 November 2025\cite{dac2026}, before the 2026 agent workflows discussed here. TCAD and TODAES likewise report completed and accepted work, including Early Access articles. The interval from research to publication may explain why the current records do not yet show a broad agentic shift.

We also collected later arXiv papers to examine newer projects, identifying our own ART preprints separately. EDA posting practices are uneven. Comparing successive submission and preprint cohorts with the four conferences and two journals can reveal how recent research enters the publication record.

\section{Discussion: directing cheaper exploration toward useful EDA}

\subsection{Industry can investigate more of its own ideas}

\begin{figure}[t]
\centering
\resizebox{\linewidth}{!}{%
\begin{tikzpicture}[x=1cm,y=1cm,font=\sffamily\figsmall]
\path[use as bounding box] (0,-.28) rectangle (8.1,4.73);
\node[anchor=west,font=\sffamily\bfseries\figtitle,text=FDdark] at (.12,4.48) {Many investigations can improve one tool};
\node[inner sep=0] at (4.04,3.04) {\includegraphics[width=2.70cm]{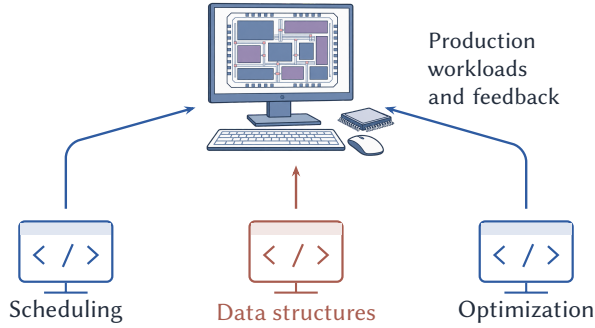}};
\node[anchor=west,align=left,text width=1.80cm] at (5.60,3.16) {Production\\workloads\\and feedback};
\pic[scale=1.25] at (1.13,.82) {terminal=Blue};
\pic[scale=1.25] at (4.04,.82) {terminal=Warm};
\pic[scale=1.25] at (6.95,.82) {terminal=Blue};
\draw[Blue,line width=.85pt,-{Stealth[length=4pt]},rounded corners=5pt] (1.13,1.38)--(1.13,2.20)--(2.77,2.66);
\draw[Warm,line width=.85pt,-{Stealth[length=4pt]}] (4.04,1.38)--(4.04,1.93);
\draw[Blue,line width=.85pt,-{Stealth[length=4pt]},rounded corners=5pt] (6.95,1.38)--(6.95,2.15)--(5.16,2.66);
\node[font=\sffamily\figtext] at (1.13,.08) {Scheduling};
\node[font=\sffamily\figtext,text=Warm] at (4.04,.08) {Data structures};
\node[font=\sffamily\figtext] at (6.95,.08) {Optimization};
\end{tikzpicture}}
\caption{The industrial opportunity includes improvements within existing tools. Cheaper exploration makes more investigations into scheduling, data structures, and optimization worth attempting against production workloads. Small gains can be reused across many designs.}
\Description{Three algorithm-development workstations feed improvements into a shared industrial EDA tool with production feedback.}
\label{fig:cost}
\end{figure}

Cheaper algorithmic exploration expands the set of ideas that a tool team can afford to investigate. Schedulers, decompositions, data structures, and local optimization choices often receive less attention than their possible value warrants. Agents can help explore these alternatives against the workloads on which the tool will actually run (Fig.~\ref{fig:cost}). Small improvements in a frequently executed kernel can matter across many designs, even when they do not constitute a new flagship algorithm.

AlphaEvolve offers an example outside EDA: its authors report a deployed scheduling heuristic that recovered an average 0.7\% of fleet-wide compute resources and an optimization effort shortened from months to days\cite{alphaevolve}. EDA tool developers already possess much of the environment needed for comparable investigations: implementations, internal netlists, and production evaluators. Integration and confidentiality still require work, but the exploration can take place close to the people who will use its results.

This affects a familiar division of labor. An industrial team might previously have left an alternative to academic researchers because investigating it required too much specialist time. When that cost falls, the team can pursue more alternatives internally. Economic accounts of AI and invention\cite{cockburn}, increasing specialization\cite{jones}, and rising research effort in technology\cite{bloom} provide a useful context for this shift.

The opportunity also gives academic work a practical direction. Public evaluators and representative workloads let smaller organizations benefit from exploration that would otherwise remain inside large tool owners. Algorithms that transfer across designs and implementations can save many teams from repeating the same investigation. EDA has a history of such shared contributions; agents can increase how extensively they are used.

\subsection{Open question: what should academic EDA contribute?}

Our trials call into question a familiar justification for academic EDA: undertaking algorithmic exploration that tool developers cannot afford to pursue themselves. If agents make that exploration inexpensive, a new method for an established formulation may provide less additional value to a tool owner, even when the improvement is real. What then justifies concentrating academic effort on another such method?

A stronger justification would be to enable work beyond a tool owner's routine search. Possible contributions include:
\begin{keypoints}
\item \textbf{A better problem or evaluator.} Expose an industrial constraint, a misleading proxy, or a failure missing from current benchmarks. This changes which improvements are worth pursuing.
\item \textbf{Transferable understanding.} Explain when a method works, identify a limitation, or provide a construction that avoids repeated search in other settings.
\item \textbf{A usable capability.} Deliver gains that survive integration and representative workloads, with enough of the implementation available for others to benefit.
\end{keypoints}

Open systems such as ABC, OpenROAD, and DREAMPlace illustrate this wider contribution\cite{abc,openroad,dreamplace}. They enable investigations beyond their authors' own applications. Such examples suggest placing more weight on shared capability and transfer across workloads. Whether these contributions can sustain a distinctive academic role as agents improve remains open; novelty within a benchmark alone offers a weaker justification.

Formulation novelty belongs in the same discussion. Our corpus distinguishes a new computational formulation from an established one, but both can provide an executable research loop. Changing an objective or adding a constraint may be useful, and agents increasingly participate in proposing such changes\cite{si}. The academic contribution is stronger when the work also establishes why the change matters. The remapping trial illustrates the issue: exact optimization of a surrogate did not ensure a better physical result. Connecting a formulation to design needs can be more consequential than another improvement against an inherited proxy.

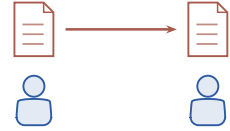
\begin{figure*}[t]
\centering
\resizebox{\textwidth}{!}{%
\begin{tikzpicture}[x=1cm,y=1cm,font=\sffamily\figtext,text=Ink]
\path[use as bounding box] (0,0) rectangle (17.4,5.9);
\draw[Rule] (8.65,.15)--(8.65,5.75) (.10,3.00)--(17.25,3.00);
\node[anchor=west,font=\sffamily\bfseries\figtitle,text=FDdark] at (.12,5.54) {a\quad Problem coverage};
\node[anchor=west,font=\sffamily\bfseries\figtitle,text=Warm] at (8.92,5.54) {b\quad Repeated combinations};
\node[anchor=west,font=\sffamily\bfseries\figtitle,text=FDdark] at (.12,2.60) {c\quad Collaboration structure};
\node[anchor=west,font=\sffamily\bfseries\figtitle,text=Warm] at (8.92,2.60) {d\quad Diffusion between groups};
\pic[scale=1.35] at (.94,4.37) {chip=Blue};
\pic[scale=1.35] at (2.74,4.37) {terminal=Warm};
\node[font=\sffamily\figsmall] at (.94,3.58) {Design};
\node[font=\sffamily\figsmall] at (2.74,3.58) {Methods};
\node[anchor=west,align=left,text width=4.35cm] at (3.93,4.25)
 {Counts and shares by problem.\\Which research loops expand?};
\foreach \y in {4.69,3.80}{
 \draw[FDlight,fill=white,rounded corners=3pt] (9.10,\y-.31) rectangle (12.53,\y+.31);
 \fill[Blue!10,rounded corners=2pt] (9.23,\y-.21) rectangle (10.59,\y+.21);
 \fill[Warm!10,rounded corners=2pt] (10.73,\y-.21) rectangle (12.40,\y+.21);
 \node[text=FDdark,font=\sffamily\figsmall] at (9.91,\y) {Method};
 \node[text=Warm,font=\sffamily\figsmall] at (11.56,\y) {Benchmark};
}
\node[anchor=west,align=left,text width=4.10cm] at (12.96,4.25)
 {Recurring combinations\\within each problem.};
\draw[Blue,line width=.85pt] (1.11,1.96)--(2.69,1.40) (1.11,.81)--(2.69,1.40);
\pic at (.91,1.96) {person=Blue};
\pic at (.91,.81) {person=Warm};
\pic[scale=1.20] at (2.87,1.40) {person=Blue};
\node[anchor=west,align=left,text width=4.35cm] at (3.93,1.40)
 {First and last authors.\\New partners or\\repeated pairs?};
\pic[scale=1.05] at (9.63,1.68) {page=Warm};
\pic[scale=1.05] at (11.88,1.68) {page=Warm};
\draw[Warm,line width=.95pt,-{Stealth[length=4pt]}] (10.04,1.68)--(11.48,1.68);
\pic[scale=.8] at (9.63,.72) {person=Blue};
\pic[scale=.8] at (11.88,.72) {person=Blue};
\node[anchor=west,align=left,text width=4.10cm] at (12.96,1.40)
 {Time until a combination\\reaches another group.};
\end{tikzpicture}}
\caption{Public-data observations for subsequent EDA cohorts. Track coverage, recurring methods and benchmarks, collaboration, and diffusion together; link preprints to their later venue versions.}
\Description{Four panels illustrate problem coverage, repeated method-benchmark pairs, first-last author collaboration, and the spread of research combinations between groups.}
\label{fig:signals}
\end{figure*}

\subsection{Open question: how should research ability be assessed?}

The trials allowed some students to engage with substantial proofs and implementations before completing the usual research-to-publication apprenticeship. That is an educational opportunity: students can encounter real research questions earlier and inspect a broader range of attempted solutions. It also changes what a completed manuscript tells a supervisor about the student's understanding.

What evidence should hiring, promotion, and funding bodies use when a paper no longer reveals how much independent judgment its production required? Reconstructing an argument, explaining a failed approach, or demonstrating a tool's sustained use are possible sources of evidence. How to turn them into a fair, scalable assessment mechanism, rather than another countable credential, is an open question.

Our own view of the necessary human contribution changed over these months. Literature reconstruction, mathematical criticism, and movement between algorithmic domains increasingly entered the delegated workflow. Harris discusses the institutional consequences of abundant technically sound research\cite{harris}; Korinek examines research automation and judgment\cite{korinek2025}; Schwartz describes a changed personal research reach\cite{schwartz}. These accounts concern other fields, but they resemble parts of the EDA experience.

Choosing questions and checking results still mattered in these trials, but their allocation changed as the tools improved. We therefore also ask which forms of judgment an assessment system should reward, and how it should adapt as agents acquire them. A mechanism built around today's division of labor may age as quickly as the capabilities it is meant to assess.

\subsection{Open questions: publication and peer review}

Our trials left us asking how human researchers can check and absorb a growing body of agent-generated results. A plausible argument may take hours to reconstruct; establishing what it adds can require still more reading and comparison. When the production of new results outpaces that work, the limiting resource becomes the attention needed to turn individual findings into shared knowledge.

Mathematics offers an early example. First Proof's second round assessed four systems on ten research problems\cite{firstproof}. Harvard's report describes a two-day refereeing workshop involving 30 mathematicians and notes that some solutions took hours to decipher\cite{mathnews}. Tao's discussion of proof abundance distinguishes producing results from verifying, understanding, and incorporating them into a field; he warns that these stages can advance at very different rates\cite{taomath}. Correct results, too, can accumulate faster than a community can evaluate and use them.

EDA conferences and journals already report rising submissions and difficulty recruiting reviewers\cite{ceda,iccad2025,aspdac2026}, alongside public concerns about review consistency\cite{dacdiscussion}. As agentic exploration develops, upcoming review cycles, possibly including DAC 2027, may begin to encounter a different balance between the cost of producing and evaluating research. We see three open questions for our field:
\begin{keypoints}
\item \textbf{Review capacity.} How can human reviewers establish correctness, novelty, and practical significance when specialist attention cannot grow as quickly as the number of results?
\item \textbf{Publishing model.} Does paper-by-paper peer review with discrete acceptance decisions remain an effective way for conferences and journals to validate and communicate rapidly accumulating knowledge?
\item \textbf{Publication-based evaluation.} If acceptance becomes a weaker proxy for research ability or practical contribution, what role should venue prestige play in evaluating people and institutions?
\end{keypoints}

Automation also reaches the reviewing side: The AI Scientist reached a conventional workshop acceptance threshold\cite{aiscientist}, and LLM assistance is already visible in review writing\cite{liangreview}. Review agents may help with some checks, but then an acceptance decision depends increasingly on how those systems are evaluated. EDA's executable tests can establish many technical properties; choosing a useful formulation or an informative comparison remains a separate task. We want to know which publishing and assessment practices can sustain those judgments at the new scale, and give appropriate credit to contributions that improve design research and tools.

\subsection{Follow changes in the research distribution}

We use the publication data as an early baseline for following this transition. Most of the conference research in the snapshot predates the agent workflows studied here. A lack of a coherent field-wide signal in that population leaves an open question for subsequent cohorts. It does not reduce the value of tracking how the distribution changes.

Counts should be read alongside coverage and concentration. Within each EDA problem, do additional papers explore more constraints and workloads, or repeat a narrower set of method--benchmark combinations? Messeri and Crockett discuss scientific monocultures associated with AI\cite{messeri}; Hao et al. distinguish individual gains from changes in collective scientific focus\cite{hao}. Work on attention in large literatures provides another reason to examine what gets reused as output expands\cite{chu}. Conversely, systems designed to seek less-attended questions illustrate how search could broaden the agenda\cite{sourati}.

These observations are possible with public records (Fig.~\ref{fig:signals}). Compare combinations against shuffled assignments that preserve their ingredients' popularity\cite{uzzi}. Keep first- and last-author positions separate to distinguish new partnerships from repeated output within the same pairs. Link preprints to later venue versions, and use first public or submission dates where available. Changing EDA preprint habits can then be studied alongside changes in research production.

The observations are most informative together. Rising output with wider problem coverage describes a different trajectory from rising output with repeated methods and benchmarks. Faster diffusion between otherwise separate groups is another change worth following. Such patterns can help the community see where cheaper exploration is opening new work and where shared evaluators, broader workloads, or a different choice of questions would be useful.

\section{Conclusion}

We deliberately explored unfamiliar EDA questions and found that substantial algorithmic work could proceed with little continuous specialist input. Some results became useful tools; others failed their practical comparison. Cheaper exploration offers tool developers more opportunities to improve their systems. For academic EDA, it also raises an open question: what should we contribute beyond increasingly inexpensive algorithm search, and how should publication, hiring, promotion, and funding recognize that value?

\section*{Data and research materials}

We assembled the paper-level corpus and semantic labels, author-role summaries, tag analyses, figure data and reproduction scripts in the companion materials, together with the ART project reports and supplied research archives. Process data identify the timestamps, ledger categories, and selected intervention episodes used here; the weekly human-time estimate is retrospective. The analyses preserve the three closure categories, distinguish primary and expanded core EDA, exclude Late Breaking Results, and retain accepted Early Access papers. The manuscript snapshot is 7 October 2026. Seven associated ART manuscripts are available as preprints\cite{routing,logic,bbyt,coarsen,clock,remap,fapo}.

\section*{AI use disclosure}

We used multiple versions of OpenAI GPT models, Kimi, and DeepSeek in the research and preparation of this article. All manuscript prose was generated by AI. Humans provided the initial initiative, checked the material, and adjusted the overall direction; the trial design sought to minimize human intervention in the specific algorithms, with the interventions described in the text. AI was also used for literature retrieval, semantic annotation, analysis code, and figure preparation. Human authors are responsible for the article's claims and citations.


\end{document}